# In-Plane and Out-of-Plane Charge Dynamics of High-$T_c$ Cuprates [★]

D. van der Marel and Jae H. Kim

*Laboratory of Solid State Physics, Materials Science Center,
University of Groningen, Nijenborgh 4, 9747 AG Groningen, The Netherlands*

**Abstract**

We propose a theoretical expression for the $k$- and $\omega$-dependent dielectric function of a stack of two-dimensional layers coupled along the direction perpendicular to the layers, and discuss some of its properties. We argue that the plasma frequencies at $k = 0$ should correspond to those which are experimentally obtained from optical measurements on *e.g.* $La_{1-x}Sr_xCuO_4$ via the $f$-sum rule analysis, regardless of the fact that such systems are strongly correlated. We discuss some of the ramifications due to strong anisotropy of the charge transport in these systems, and the lack of coherence for the transport in the direction perpendicular to the layers.

In theoretical discussions of the high-$T_c$ cuprate superconductors it is usually emphasized, that there exists a strong anisotropy in the electrodynamical properties between the in-plane and out-of-plane directions. Usually this anisotropy is regarded as resulting from a large mass anisotropy of the charge carriers, so that effectively $m_c \gg m_{ab}$ where $m_c$ and $m_{ab}$ are the in-plane and out-of-plane effective masses. Ideally one prefers to use a model dielectric function where these superconductors are treated as a stack of layers which are completely decoupled in the $c$ direction, which amounts to taking $m_c \to \infty$. Such models are indicated as layered electron gas (LEG) models. An important consequence $m_c \to \infty$ is, that the plasmon dispersion in the long-wavelength limit ($kd \ll 1$ where $k$ is the wavevector and $d$ is the interlayer spacing) is given by $\omega(k)/\omega_{ab} \approx k_{ab}/\sqrt{k_{ab}^2 + k_c^2}$. For a given finite value of $k_c$ this is an acoustical dispersion as a function of $k_{ab}$. If we write the dispersion in the form $\omega(k)/\omega_{ab} \approx \sin\theta$, where $\theta$ is the angle between $\vec{k}$ and the $c$ axis, it becomes clear, that $\vec{k} \to 0$ is a singular point in the dispersion, with $\omega_c = 0$ for $\vec{k} \parallel \vec{c}$, whereas for $\vec{k} \perp \vec{c}$ the plasma frequency is $\omega_{ab} = (4\pi n e^2/m_{ab})^{1/2}$.

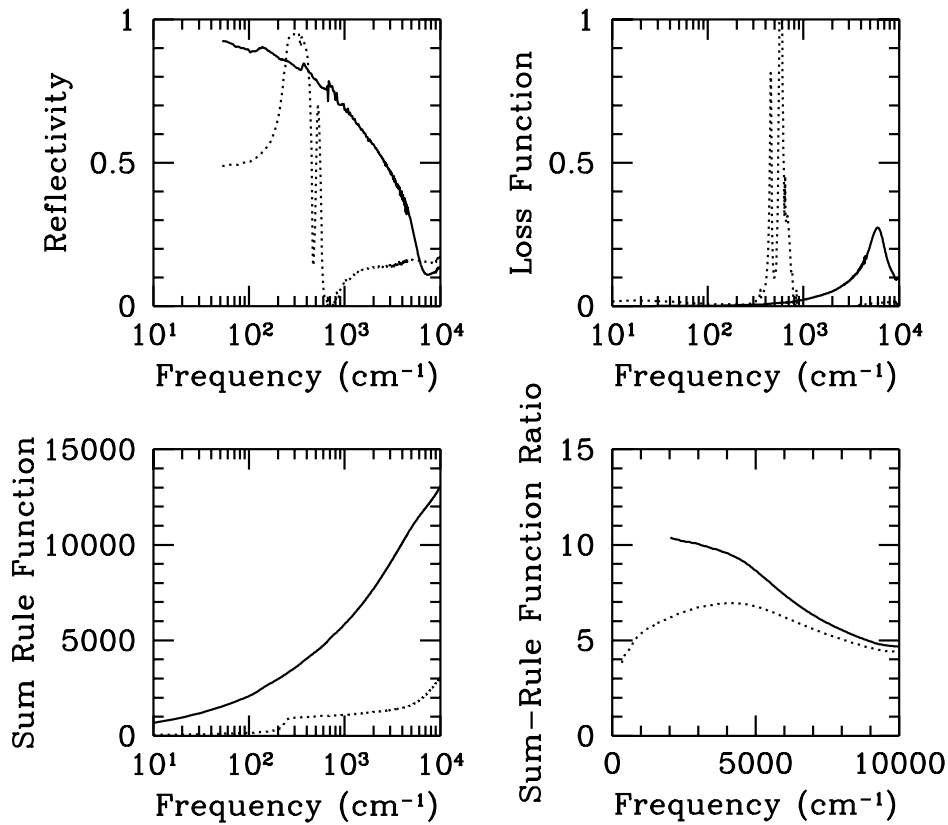

Fig. 1. Spectral functions of $La_{1.9}Sr_{0.1}CuO_4$ at room temperature for $ab$ plane (solid curves except (d)) and $c$ axis (dotted curves except (d)). In (d) we display $\omega_{ab}/\omega(c)$ obtained from Eq. (1) with (solid) and without (dotted) correction for the phonon-contributions.

Indeed it is known from optical measurements on cuprates, that with the electric field oriented along the planes a plasmon is observed at around 1 eV [1–3] whereas above $T_c$ there is no zero crossing of $\mathrm{Re}\epsilon_c$ associated with a free-carrier plasmon (only optical phonons are observed)[4–7]. In Fig. 1 we demonstrate this by plotting the energy loss function ($-\mathrm{Im}1/\epsilon$) for both directions as obtained from our optical experiments. However, the actual situation is more complicated in two respects:

(1) In the normal state the absence of the plasmon along the $c$ direction can *not* be attributed to having $m_c \to \infty$, but is rather due to overdamping. This can be seen in the following way: With finite damping $\epsilon = \epsilon_\infty - \omega_p^2/\omega(\omega + i\gamma)$ is the (optical) dielectric function at $k \approx 0$, so that for $4\omega_p^2 < \epsilon_\infty \gamma^2$ the zero crossings are along the *imaginary* frequency axis, i.e. for $2i\omega = \gamma \pm \{\gamma^2 - 4\omega_p^2/\epsilon_\infty\}^{1/2}$.

On the other hand, one can use the *f*-sum rule

$$\int_0^x \omega \text{Im} \epsilon(\omega) d\omega = \frac{\pi}{2} \omega_p^2 \quad (1)$$

to obtain $\omega_c$ from the optical data by integrating the imaginary part of the dielectric function upto a suitably chosen cutoff frequency $x$ below the onset of interband transitions (in practice such a clean separation is not always possible). In Fig. 1(c) we display $\omega_p$ as obtained from Eq. (1) as a function of cutoff frequency $x$ along the horizontal axis. The step at 200 cm$^{-1}$ in the *c*-axis data, as well as a somewhat broader step at 150 cm$^{-1}$ in the *ab*-plane data are phononic contributions. After correction for these phonon-contributions we find that the ratio of *electronic* plasma frequencies $\omega_{ab}/\omega_c$ drops from 10 to 4 as a function of increasing cutoff frequency. The ratios with (solid line) and without (dashed curve) correction for the phonon contributions are displayed in Fig. 1(d). From LDA band structure calculations we obtained $\omega_{ab}/\omega_c = 5$, in good agreement with the experimental sum-rule analysis. By fitting the optical data to a Drude model, we obtained for the optical scattering-rate that $\tau_{ab}/\tau_c > 50$, which is consistent with the DC conductivity ratio.

(2) In the superconducting state there is a *partial* recovery of a plasmon in the *c* direction, essentially because there exists a finite probability for pair-hopping in the superconducting state. This is equivalent to having a small but finite critical current perpendicular to the layers, which has a finite penetration depth associated with it. Using again sum rule arguments, we expect that the penetration depth is now given by the dirty limit formula $2\pi\lambda_c = c(3E_g\sigma_c(0)/\hbar)^{-1/2}$, which is determined by the 'missing area' in the optical conductivity. The zero crossing of Re$\epsilon$ occurs now at frequencies below the superconducting gap, and corresponds to a collective mode of the superconducting condensate at a frequency $\tilde{\omega}_\phi^2 = c^2/\lambda_c^2\epsilon_\infty$. The ratio between this frequency and the undamped normal-state plasmon in the *ab* direction is

$$\frac{\tilde{\omega}_\phi}{\omega_{ab}} = \left[\frac{\epsilon_{ab\infty} m_{ab} \arctan(2.3 E_g/\hbar\gamma)}{\epsilon_{c\infty} m_c \pi/2}\right]^{1/2}. \quad (2)$$

The left side of the expression is $50/12000 = 0.009$, whereas the righthand side becomes 0.011, assuming that $E_g = 3.5 k_B T_c$ and using the other parameters as discussed above. Hence if we take into account that $m_c/m_{ab} < 40$ and that $\tau_{ab}/\tau_c > 50$, we obtain a consistent description of the anisotropies in $\sigma(0)$, the penetration depth, the screened plasma frequency, and the bare plasma frequency obtained from the *f*-sum rule.

From the above discussion, we may conclude that the LEG model for the anisotropic dielectric properties of the high-$T_c$ cuprates is not realistic. The major flaw is the fact that taking $m_c \to \infty$ is a gross oversimplification. The second problem arises due to the strong damping terms. Microscopically the anisotropic damping is quite interesting, and has been predicted as a result of spin-charge separation [8,9]. To derive the full $k$- and $\omega$-dependent dielectric function including the damping effects is a very difficult theoretical problem. The presence of strong anisotropy makes the solution of this problem already quite formidable even if the only source of scattering is (elastic) impurity scattering, which becomes more difficult if the damping has a many-body origin. In this paper we present a formula for the $k$- and $\omega$- dependent dielectric function without taking into account damping. Although there is no hope that such an expression for $\epsilon$ gives a complete description of the optical properties or electron-energy loss function of high-$T_c$ cuprates, the expressions are nevertheless quite relevant in the discussion of the optical sum rule. Even though the optical spectral shape will be very different from that of the undamped and unrenormalized plasma poles discussed above, the imaginary part of the dielectric function should still obey Eq. (1).

Let us extend the a layered electron gas model[10,11], by including a finite hopping between the planes. The electrons are confined to a stack of 2D sheets a distance $l_c$ apart, coupled by a finite interlayer hopping parameter $t$. The dispersion relation close to the Fermi-level is

$$E_k = E_F + v_F\hbar(k_\| - k_F) + 2t\cos(k_\perp l_c) \qquad (3)$$

We indicate such systems as a coupled layered electron gas (CLEG). The plasmon-dispersion was derived for such a model by Grecu[12] using perturbation theory. The bare polarization propagator for $k_B T \to 0$ and $\vec{q} \to 0$ can be expressed as an integral over the Fermi surface

$$\Pi^0(\vec{q},\omega) = 2\frac{\Omega_u}{8\pi^3}\int_{S_F}\frac{\vec{q}\cdot d\vec{s}}{\hbar\omega - \hbar\vec{q}\cdot\vec{v}(\vec{s})} \qquad (4)$$

Using straightforward mathematical manipulations, the full expression for the k and $\omega$ dependent polarizability can be easily obtained. To simplify the expression we introduce the effective velocity perpendicular to the layers $v_c \equiv 2\hbar^{-1}tl_c$. In calculating $\Pi^0$ we use the fact that the angular averages of terms of the type $\cos\theta\sin^{2n+1}\theta$ are all zero, so that we obtain

$$\Pi^0(\vec{q},\omega) = 2\frac{k_F\Omega_u}{4\pi^3\hbar v_F l_c}\int_0^{2\pi}d\theta\int_0^{2\pi}d\phi\frac{v_F q_\| \cos\phi - v_c q_\perp \sin\theta}{\omega - (v_F q_\| \cos\phi - v_c q_\perp \sin\theta)} \qquad (5)$$

In reciprocal space the expression for the Coulomb interaction between electrons confined to an infinite stack of $\delta$-layers is $v_q = 4\pi e^2 S|\vec{q}|^{-2}$ where $S$ is the form-factor of the single electron wavefunctions. For $\delta$-layers this is

$S(\vec{q}l_c) = S(\vec{x}) = \frac{1}{2}x_\|^{-1}|\vec{x}|^2 \sinh(x_\|)/\{\cosh(x_\|) - \cos(x_\perp)\}$. The dielectric function as obtained from the random phase approximation is $\epsilon = 1 - v_q \Pi^0$. In the limit $\vec{q} \to 0$ we obtain the plasma frequencies $\omega_{p\|} = e(2k_F v_F)^{1/2}(\hbar l_c)^{-1/2}$ and $\omega_{p\perp} = 2et(2k_F l_c)^{1/2}\hbar^{-3/2}v_F^{-1/2}$ for the directions of propagation parallel and perpendicular to the planes. In the case of LEG the dielectric function is

$$\epsilon(\vec{q},\omega) = 1 - 2S \left[\frac{\omega_p}{|\vec{q}|v_F}\right]^2 \left(\left(1 - \frac{q_\|^2 v_F^2}{\omega^2}\right)^{-1/2} - 1\right) \quad (6)$$

On the other hand, from LDA band structure calculations we obtained, that for doped $La_2CuO_4$ the ratio $\omega_{p\perp}/\omega_{p\|} \approx 0.2$, from which we conclude that the full plasmon-dispersion formula should be used. The CLEG dielectric function using the RPA approximation is

$$\epsilon(\omega,\vec{q}) = 1 - \left[\frac{q_\|}{|\vec{q}|}\right]^2 \frac{S\omega_{p\|}^2 p^0\left(\frac{v_F q_\|}{\omega}, \frac{v_c q_\perp}{\omega}\right)}{\omega(\omega+i0^+)} - \left[\frac{q_\perp}{|\vec{q}|}\right]^2 \frac{S\omega_{p\perp}^2 p^0\left(\frac{v_c q_\perp}{\omega}, \frac{v_F q_\|}{\omega}\right)}{\omega(\omega+i0^+)} \quad (7)$$

where $p^0$ is the function

$$p^0(a,b) \equiv \frac{2}{\pi a} \int_0^\pi \{(1 - a\cos\phi)^2 - b^2\}^{-1/2} \cos\phi \, d\phi \quad (8)$$

Let us first consider the plasma dispersion. In Fig. 2 we display this dispersion for the LEG and the CLEG models. The dispersion of the upper branch calculated with these parameters is in agreement with electron energy loss experiments of high-$T_c$ cuprates[13,14]. The c-axis plasma frequency is the lower bound of a continuum of plasmon states. In the limit $\vec{q} \to 0$ both $S(\vec{q}l_c) \to 1$ and $p^0 \to 1$, and the optical conductivity for $\vec{E}$ parallel and perpendicular to the planes becomes $\sigma_{jj} = \omega_{pj}^2/4\pi(\gamma_j - i\omega)$.

Let us first point out that the above formula for $\epsilon$ contains two nontrivial terms. In the small-$\vec{q}$ limit (where the distinction between longitudinal and the transverse dielectric functions disappears), the first term describes the dielectric function for fields parallel to the planes, whereas the second one enters the description for electric fields perpendicular to the planes. Although we have not been able to give a microscopic justification hereof, the experimental results [4,5,15,1] indicate that the many-body effects affect these two terms in a very different way: in the cuprates there is a strong anisotropy (of the order of hundred), both in $\omega_{pj}^2$ and in the scattering rate $\gamma_j$. In fact, the c-axis conductivity turns out to be practically frequency-*in*dependent, whereas the in-plane conductivity has a relatively small (but frequency dependent) scattering rate. These phenomena cannot be explained from weak elastic or inelastic scattering of fermionic charge carriers [16,17], but rather require charge carriers with a renormalization factor $Z$ which is very small or zero. Anderson and coworkers address all these issues within a single framework of carrier confinement to the planes due to the formation of a 2D Luttinger liquid, with a de-confinement due to the onset of superconductivity [8,9].

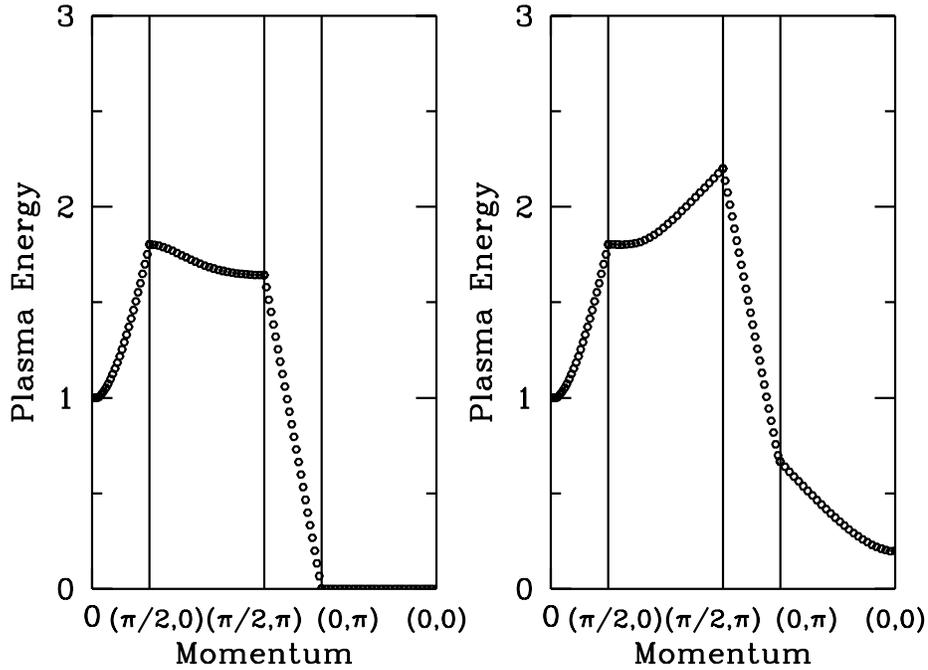

Fig. 2. Plasmon dispersion of a layered electron gas (left) and that of a coupled layered electron gas with $\omega_\perp/\omega_\parallel = 0.2$. The reciprocal-space coordinates are indicated as $[q_\parallel l_c, q_\perp l_c]$ at a number of points. For the Fermi velocity we assumed $v_F/l_c\omega_\parallel = 1$.

As we discussed above, the effect of finite scattering is a quite formidable theoretical problem yet to be solved. However, from inspection of Eq. (7), and Fig. 2, we see that the first term determines the dispersion of the plasma frequency with $q_\parallel$, whereas the dispersion with $q_\perp$ due to the second term is relatively small. A simple, and somewhat crude 'shortcut', which is only valid for the collective-mode part of the dielectric function (*i.e.* for frequencies outside the particle hole continuum), is to replace each of the two terms with a $q$-dependent plasma pole, each of which has a different damping term. This suggests that we can make the following approximation in the normal state

$$\epsilon(\omega, \vec{q}) = 1 - \frac{q_\parallel^2}{|\vec{q}|^2} \frac{\omega_p(q_\parallel)^2}{\omega[\omega + i\Gamma_s(\vec{q},\omega)]} + \frac{q_\perp^2}{|\vec{q}|^2} \frac{4\pi i \sigma_s(\omega)}{\omega}. \tag{9}$$

This approximation is consistent with our experimental results at $k$=0. The modifications of this function in the superconducting state are quite small, and require a very thorough analysis, as in *e.g.* [18,19]. This dispersion and the corresponding loss functions could be measured with *e.g.* electron energy loss spectroscopy or with inelastic X-ray scattering.

**Conclusions**

From an analysis based on the *f*-sum rule we showed, that the experimental values for the anisotropy in plasma frequency along and perpendicular to the planes is between 4 and 10. We conclude that the absence of a *c*-axis plasmon in the normal state of high-$T_c$ cuprate superconductors results from a non-Drude smearing of the optical conductivity perpendicular to the $CuO_2$ planes over a very wide energy range. This assignment alleviates the discrepancy by two orders of magnitude existing in the literature, between the plasma frequency obtained from optical experiments and that from the single-electron interlayer-hopping matrix element. We present an analytical expression for the $k$ and $\omega$ dependent dielectric function for an electron gas, with a Fermi surface in the shape of a corrugated cylinder. We propose a phenomenological extension of this expression, to take into account the incoherent nature of the charge transport in the *c* direction.

**Acknowledgments** We gratefully acknowledge Ir. H. S. Somal for his assistance with the experiments, and V.H.M. Duijn, N.T. Hien and A.A. Menovsky for providing single crystals. This investigation was supported by the Netherlands Foundation for Fundamental Research on Matter (FOM) with financial aid from the Nederlandse Organisatie voor Wetenschappelijk Onderzoek (NWO).